\documentclass[12pt]{article}
\usepackage{epsfig,textpos,url}
\usepackage{lineno}

\begin{document}
\title{Controlling a digital twin\\ of the CEBAF injector from EPICS}
\author{Jude Doredant, Georgia College and State University\\
  Volker Ziemann, Jefferson Lab}
\date{August 25, 2026}
\maketitle
\begin{abstract}\noindent
  The Experimental and Instrumental Control System EPICS is used to control magnets
  and to read data from diagnostic devices in the CEBAF accelerator. We describe
  a prototype system where an EPICS installation, local on a laptop or Raspberry Pi,
  controls beam optics simulations. They can be running on the same computer or on
  a separate computer connected via the network. The prototype focuses on modeling
  and controlling the moderately complex injector beam line of CEBAF, which only
  comprises a few dozen magnets and beam monitors.
\end{abstract}
%
%
\section{Introduction}
Digital twins of particle accelerators offer great opportunities to
improve their performance~\cite{TWINAC,TWINAC2,KAM,LUME,BHARDWAJ,KHAIL},
provide wonderful test beds to train Artificial Intelligence and Machine
Learning (AI/ML) tools~\cite{KREUZ}, and will help to train new staff without
adversely affecting the hardware of an accelerator serving experiments.
\par
In this report, which evolved from a summer-student project report~\cite{JUDE},
we mostly have the latter aspect in mind and develop a digital twin of the
injector  of the Continuous Electron Beam Accelerator Facility (CEBAF)
at Jefferson Lab. The twin itself is written in Matlab~\cite{MATLAB} and
is based on the software from~\cite{VZAPB}. We then enhanced the software
to listen to a network socket, from where it receives commands that
resemble the Standard Commands for Programmable Instruments
(SCPI)~\cite{SCPI}. This type of protocol interfaces easily with the
StreamDevice library~\cite{STREAM}, which is an add-on to the Experimental
Physics and Industrial Control System (EPICS)~\cite{EPICS}. We can either
use an existing EPICS installation or following the instructions
from~\cite{VZHOS}, install the software on another computer. In either
we prepare the configuration files for EPICS and start Input-Output
Controller (IOC). On the one hand, the IOC communicates with the twin
and on the other hand, it publishes so-called EPICS process variables
(PV) on the network that can be used by any EPICS-aware program, including
those that normally communicate with the hardware. Only in this case,
they control the twin. Exposing the internal simulation parameters as
EPICS process variables resembles the concept pursued in~\cite{LUME}.
\par
In the remainder of this report, we first discuss Matlab implementation
of the digital twin of the CEBAF injector in Section~\ref{sec:twin},
followed by a description of the SCPI-based communication protocol and
how it is implemented in the twin. Section~\ref{sec:epics} discusses how to
configure the EPICS configuration files that make it possible to seamlessly
communicate between the Matlab twin and EPICS. Finally, Section~\ref{sec:use}
presents examples of how the system can be used to train colleagues
unfamiliar with the CEBAF injector. Conclusions round up our presentation.
%
\section{Digital Twin in Matlab}
\label{sec:twin}
We chose Matlab and the software from~\cite{VZAPB} as the basis for
the twin, because it gives full control of every aspect of the simulation
engine and thereby provides a great learning experience. Yet, however
rudimentary it is, it adequately describes all the linear beam optics,
including misalignment of components. Moreover, new components, such
as Wien filters~\cite{WIEN}, are easily added.
\par
We describe the sequence of elements in the following tabular way,
which is a extension from the one used in~\cite{VZAPB}
\begin{verbatim}
   code  repeat  length  strength  name  extra1  extra2
\end{verbatim}
where {\tt code} characterizes the type of element, {\tt repeat}
defines how many times the element is repeated, and {\tt length},
{\tt strength}, and {\tt name} are self-explanatory, where {\tt name}
agrees with the name used in the real accelerator. The two extra
parameter are used to describe the transverse displacement of
focusing magnets, and kick angles for steering magnets. In the
Matlab implementation the data are thus stored in an array with
one row for every element and six columns for the six parameters
{\tt code}, {\tt repeat}, {\tt length}, {\tt strength}, {\tt extra1},
and {\tt extra2}. The main purpose of the twin is then changing
these six parameters and recalculating the beam optics, including
beam sizes, beta functions, the the trajectory.
\par
After reading this input file, the software calculates the transfer
matrices, beta functions, and the trajectory. The basic software is
described in detail in~\cite{VZAPB}. It takes the four-dimensional
coupled optics into account, though extending it to include the
longitudinal phase-space variables is straightforward. Compared to
the code from~\cite{VZAPB}, in the twin, we additionally take the
misalignment of elements and the kicks from the steering magnets
into account when calculating the trajectory.
\par
\begin{figure}[p]
\begin{center}
  \includegraphics[width=0.9\textwidth]{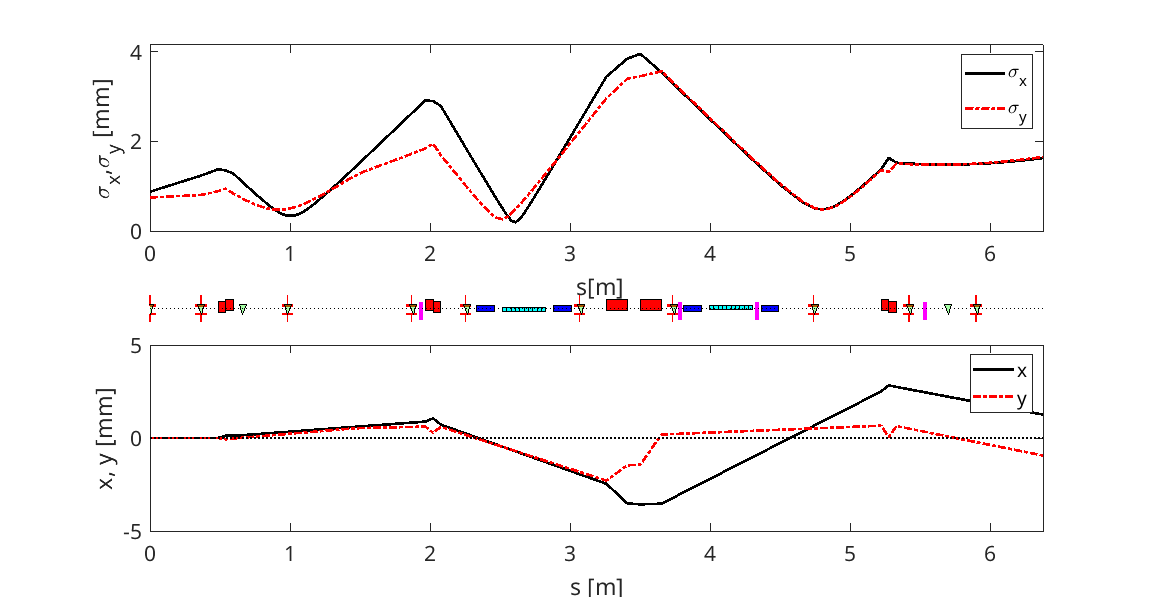}
\end{center}
\caption{\label{fig:dt1}Beta functions (top) and beam trajectory (bottom)
  generated inside Matlab to display the ``real'' state of injector.}  
\begin{center}
  \includegraphics[width=0.9\textwidth]{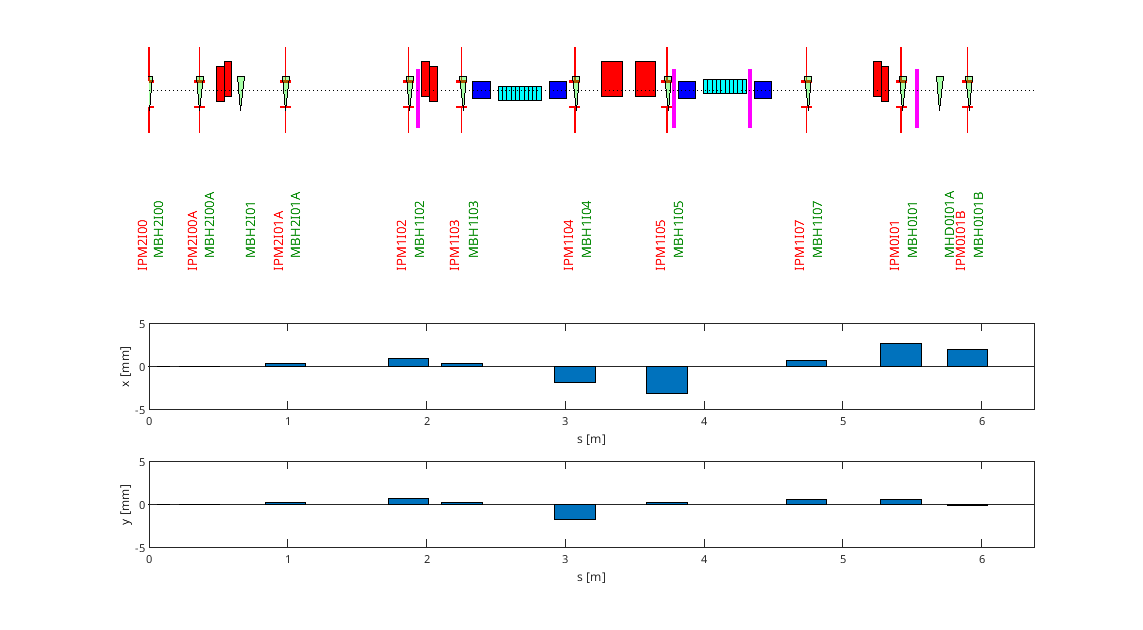}
\end{center}
\caption{\label{fig:dt2}Top: beam line description with beam position
  monitors (red) and steerers (green) identified. Bottom: horizontal and
  vertical beam position monitor signals.}
\end{figure}
The communication with the outside is implemented by first opening a
server socket and then wrapping the beam optics and trajectory calculations
in an infinite loop that reads from the socket, parses the received command,
and updates the beam line description before jumping back to the start,
update the figures from Figure~\ref{fig:dt1} and~\ref{fig:dt2}, and wait
for a new command. The following code snippet illustrates the process
\begin{verbatim}
  s=tcpserver(8000,'TimeOut',1e20);
  configureTerminator(s,"CR/LF");
  while 1
    if update_flag==1  % optics update and new figures
      [Racc,spos,nmat,nlines,state]=calcmat(beamline,state0);
      do_all_plots;  drawnow('update');  
    end    
    line=char(s.readline); % read command from socket
    token=split(line); 
    switch length(token)   % how many arguments? 
      case 1  % request a single value
          :
      case 2  % name and value, set a parameter 
          :
\end{verbatim}
The {\tt tcpserver} opens a socket that listens on port {\tt 8000}
for new commands before we enter the infinite {\tt while 1} loop.
Inside the loop, we update the optics and the figures, if magnet
values have changed. If, for example, only position monitor readings
are requested, there is no need to update the beam optics calculations.
The command {\tt s.readline} returns the received string that we
convert to a character array and split that into tokens. If only
one token is present, it is a request to return some value. If
two tokens are present, one is the name of a parameter and the
other is the value. We then set a parameter to the new value. We
defer the description of the parsing to the next section, where
we discuss the protocol in more detail. 
\par
\begin{figure}[tb]
\begin{center}
  \includegraphics[width=0.9\textwidth]{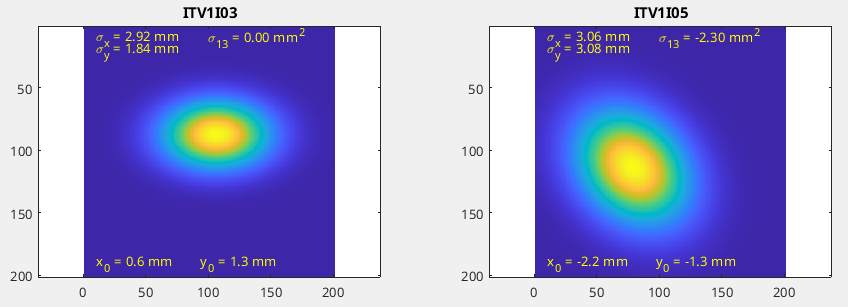}
\end{center}
\caption{\label{fig:dt3}The viewer images on the first two screens
  with annotations of beam sizes and centroid positions.}
\end{figure}
Figure~\ref{fig:dt1} is updated in the {\tt do\_all\_plots}
script and displays the projected horizontal (black) and vertical
(red) beam sizes along the beam line in the upper panel. Immediately
below, a rendition of the sequence of elements is shown. Solenoids
and quadrupoles are shown as red and blue rectangles, respectively.
If they are displaced upward, their polarity is positive, otherwise
it is negative. Beam position monitors are shown as small red
capacitors, and steering magnets as green triangles. Larger vertical
lines in cyan indicate the positions of beam viewers. The elongated
rectangles in magenta are the vertical (displaced downward) and
horizontal (displaced upward) Wien filters. The panel on the bottom
shows the horizontal (black) and vertical (red) trajectories,
respectively. As we update magnet values, the beta functions and
trajectories follow in real time.
\par
The topmost panel in Figure~\ref{fig:dt2}, also updated in
{\tt do\_all\_plots}, repeats the sequence of
elements. The panel immediately below displays the names of position
monitors (red) and steering magnets (green). The two panels at the
bottom show the positions, as recorded by the beam position monitors
(BPM), as bargraphs; the upper one for the horizontal positions and
the lower one for the vertical positions. As magnet values are changed,
the position monitor values follow in real time. 
\par
Figure~\ref{fig:dt3} shows the viewer images from the first two screens
in the injector beamline where the beam intensity is sampled at 0.1\,mm
pixel size over a range of $10\,$mm $\times 10\,$mm. Horizontal and
vertical beam sizes as well as the correlation $\sigma_{13}$ are annotated
at the top of the respective images and the centroid position is shown
near the bottom. Observe the non-zero correlation on the right-hand plot,
which stems from the so-called spin-flipper solenoids between the
Wien filters. They are not counter-wound because they have to rotate
the polarization and therefore introduced cross-plane coupling in the
betatron motion. We point out that the plots from Figure~\ref{fig:dt1},
\ref{fig:dt2}, and~\ref{fig:dt3} are directly generated by the Matlab
twin.
\par
In the next section we now turn our attention to the parser of the
commands received via the socket. It is tightly linked to the protocol
that we use to communicate with the twin.
\section{Communication Protocol}
\label{sec:comm}
Most modern test and measurement instruments adhere to some version
of the SCPI protocol, which is based on sending text strings between
devices. Usually one can request a value of a parameter {\tt P}
by sending the name with an appended question mark, here for example
{\tt P?}. The server, in our case the twin, will then respond by
sending an echo of the parameter name and a value, for example
{\tt P 1015}, if we assume that {\tt P} represents the ambient
pressure and the value is given in milli-bar.
In order to set a value, for example the voltage {\tt V} to a
new value we simply send {\tt V 4.5} to set the voltage to 4.5\,V.
\par
We thus have to come up with a naming scheme for the parameters that
control our beam optics simulation. To do so, we follow the convention
normally used in CEBAF. Magnet strength values are characterized by
the name of the magnet with ``{\tt .BDL}'' appended. For example we
can request the excitation of the excitation of the first steering
magnet from Figure~\ref{fig:dt2} by sending {\tt MBH2I00H.BDL?}
to the twin. It will then respond by sending the value of the
{\tt extra1} parameter back to the requester. Here the character
{\tt H} is also appended to denote the excitation of the horizontal
steering magnet. Likewise, {\tt V} denotes the vertical steerer.
Setting a new value is done by sending {\tt MBH2I00H.BDL -2.1},
where presently we interpret the numerical value as the kick
angle in mrad. Adapting the system to handle magnet excitations
in $\mu$Tm is simply a matter of dividing by the momentum, a feature
that we will add in the near future. Other magnet excitations are
handled in a very similar fashion.
Beam position monitor signals are requested by sending the device
name and {\tt .XPOS} or {\tt .YPOS} with a question mark
appended. Thus the vertical position of the last BPM is requested
by sending {\tt IPM0I01B.YPOS?} to the two. It then returns
{\tt IPM0I01B.YPOS 1.1}, where {\tt 1.1} is the vertical displacement
of the beam in mm.
\par
The following code snippet illustrates the decoding of a request
to return a magnet value 
\begin{verbatim}
   switch length(token)  % how many arguments?
     case 1  % request a single value
       update_flag=0; 
       if ...
          :
       elseif contains(line,'H.BDL?')  % check HCOR
         k=strfind(line,'H.BDL?'); nam=line(1:k-1); 
         magpos=ifind(nam,names);
         s.writeline([nam,'H.BDL ',num2str(beamline(magpos(1),5))]);
       elseif ...
\end{verbatim}
Since only a single token was sent, it is a request and there is no need
to update the beam optics calculations. Instead, we locate
the substring {\tt H.BDL?}. If it is found, the name {\tt nam} of the
device is the string up to the location {\tt k} where  {\tt H.BDL?} starts.
Then we locate the line {\tt magpos} in the lattice description, and write
the fifth entry, the horizontal kick, back to the socket {\tt s}. Decoding
other requests works similarly.
\par
If on the other hand, two tokens arrived, we have to set a parameter in
the beamline description, which is illustrated in the following snippet
\begin{verbatim}
    case 2  % set a value, name+value provided
      update_flag=1;
      if contains(char(token(1)),'H.BDL')  % set HCOR
        k=strfind(line,'H.BDL'); nam=line(1:k-1); 
        magpos=ifind(nam,names);
        beamline(magpos,5)=str2double(token(2));
      elseif ...
\end{verbatim}
First we ensure that the beam optics calculations are updated and try to
locate the substring {\tt H.BDL} in the first token. If found, we
determine the magnet name and position {\tt magpos} in the same way as
before and update the fifth position in the beamline description with
the value that has arrived in the second token, but converted to a
double or float value.
\par
Presently the code of the digital twin is rather compact, it is less
than two hundred lines of Matlab, excluding plotting and beam optics
calculations. Yet, it allows us to update all beam optics calculations
by interacting via text-based commands from the outside. One particularly
attractive controller is based on the EPICS control system. 
\section{EPICS}
\label{sec:epics}
As mentioned in the introduction, an EPICS IOC is a server process that
communicates with twin and then publishes process variables on the network
such that other programs can use them. This functionality is encoded in
two types of configuration files. First, protocol files configure the
communication with the hardware; in our case with the twin. We thus have
to encode our SCPI-like protocol. For example, reading and setting the
excitation of the first steering magnet in the CEBAF injector is handled
by the following code snippet
\begin{verbatim}
  # ./dtwinApp/Db/magnets.proto
  Terminator = CR LF;
  get_MBH2I00H {
    out "MBH2I00H.BDL?";
    in "MBH2I00H.BDL %f";
    ExtraInput = Ignore;
  }
  set_MBH2I00H {
    out "MBH2I00H.BDL %f";
    ExtraInput = Ignore;
    @init { get_MBH2I00H; }
  }
          :
\end{verbatim}
Functions for other devices follow in the same file. The first line with the
prepended {\tt \#} specifies the name and location of the file and the then
we define the end of line character. The function {\tt get\_MBH2I00H}
sends the request string {\tt MBH2I00H.BDL?} and then expects an echo of the
parameter name and a float value {\tt \%f} in return, while other spurious
characters should be ignored. The function {\tt set\_MBH2I00H}, on the other
hand, only sends the parameter name and a float value {\tt MBH2I00H.BDL \%f}.
Additionally, the {\tt @init} statement ensures that the value in the twin
at startup of the IOC is read, such that the IOC always represents the current
state of the twin.
\par
The second type of configuration files are database files, which link the
protocol files to process variables. The database entry that does this for
the first corrector magnet is shown in the next code snippet
\begin{verbatim}
  # ./dtwinApp/Db/magnets.db
  record(ao, "$(USER):MBH2I00H:BDL") {
    field(DESC, "magnet MBH2I00H:BDL value")
    field(DTYP, "stream")
    field(OUT, "@magnets.proto set_MBH2I00H $(PORT)")
  }
        :
\end{verbatim}
where many additional records for other devices follow; they all use the same
template. We point out that we have to use the colon instead of a period
between name and {\tt BDL}, because the period is a protected symbol in our
basic EPICS installation.
The variable strings {\tt \$(USER)} and {\tt \$(PORT)} are assigned in the server
program that we discuss below. These strings are automatically replaced at the
time the database file is loaded. This snippet defines a analog output {\tt ao}
record with the name {\tt DT:MBH2I00H} where we assume that {\tt \$(USER)}
is specified as {\tt DT}, for {\em digital twin}. The field names {\tt DESC} gives
some human-readable  information and the field {\tt DTYP} ensures that the
record is handled by the StreamDevice library. Finally, the last line of type
{\tt OUT} links this record to the function {\tt set\_MBH2I00H} from the file
{\tt magnets.proto}. {\tt \$(PORT)} points to the physical channel through
which the information flows. This could be a serial port or a network socket;
in our case it is the server socket on the twin.
Both protocol and database files have a very simple structure and are generated
quasi-automatically from the beamline description, either using a python
script~\cite{JUDE} or from within Matlab. All we have to do is to copy them to
the correct place so that epics finds them.
\par
In order to enhance the training potential of the twin, we added a number of
EPICS commands. For example, {\tt caput DT:RESET 1} reloads the beamline description
from file and {\tt caput DT:MISALIGN 0.3} sets all quadrupole and solenoid displacement
random values  with an rms of 0.3\,mm. {\tt caput DT:SCREEN ITV1I03} saves the image
for screen named {\tt ITV1I03} to disk and fills the beam sizes and the
correlation in an EPICS process variable that can be retrieved with
{\tt caget DT:SIGMAS}.
\par
The IOC is initiated by executing a process that is typically called {\tt st.cmd}.
The key parts are reproduced in the following snippet
\begin{verbatim}
  drvAsynIPPortConfigure("SOCKET1","127.0.0.1:8000",0,0,0)

  dbLoadRecords("dtwinApp/Db/magnets.db","PORT='SOCKET1',USER='DT'")
  dbLoadRecords("dtwinApp/Db/bpm.db","PORT='SOCKET1',USER='DT'")
        :
\end{verbatim}
where the first line defines a communication channel named {\tt SOCKET1}
that points to port {\tt 8000} at IP number {\tt 127.0.0.1} which is the
{\tt localhost}, or the same computer the IOC runs on. The following two
lines load the database records for magnets and for beam position monitors.
Here also the strings for {\tt \$(USER)} and {\tt \$(PORT)} are assigned.
Once all configuration files are prepared, we compile the IOC and start
it, whereupon we can access all process variables, for example, with the
command line interfaces {\tt caget} and {\tt caput}.
\section{Using the system}
\label{sec:use}
In our first test, we continuously monitor the horizontal position on the
second BPM with {\tt camonitor DT:IPM2I00A:XPOS} and change the horizontal
angle of the upstream horizontal steering magnet by 1\,mrad with
{\tt caput DT:MBH2I00H 1}. This causes the position on the BPM to vary by
0.36\,mm, which corresponds to the distance from the steering magnet to
the BPM.
\par
\begin{figure}[tb]
\begin{center}
  \includegraphics[width=0.38\textwidth]{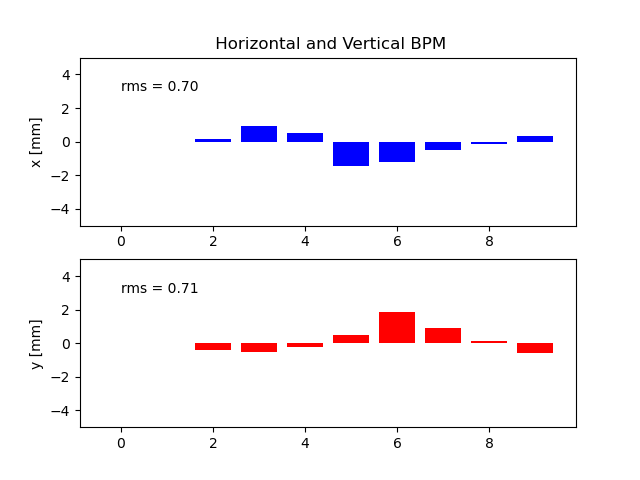}
  \includegraphics[width=0.61\textwidth]{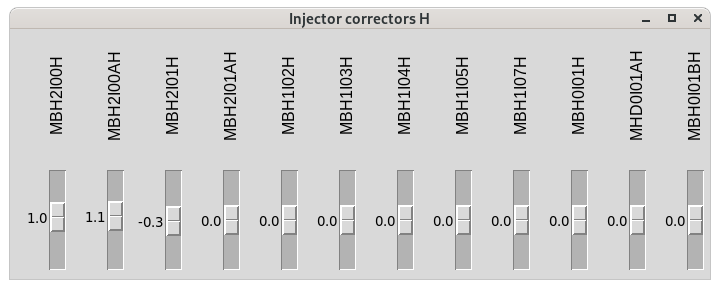}
\end{center}
\caption{\label{fig:sbpm}Display of the beam positions generated by a python
  script that reads the EPICS process variables (left) and an interface to
  set the values of the horizontal steering magnets (right).}
\end{figure}
Instead of relying on the Matlab output to show the BPM positions, we can
prepare a python script to read the process variables and display them
with {\tt matplotlib}. The shortened version of that code is shown in the
following snippet
\begin{verbatim}
         :
  bpmx=['DT:IPM2I00:XPOS','DT:IPM2I00A:XPOS','DT:IPM2I01A:XPOS',
        'DT:IPM1I02:XPOS', 'DT:IPM1I03:XPOS', 'DT:IPM1I04:XPOS',
        'DT:IPM1I05:XPOS', 'DT:IPM1I07:XPOS', 'DT:IPM0I01:XPOS',
        'DT:IPM0I01B:XPOS']
  plt.ion()
  while True:
    bpmxpos=epics.caget_many(bpmx)
    plt.clf()
    plt.bar(range(len(bpmxpos)),bpmxpos,color='blue')
         :
\end{verbatim}
After importing all needed python libraries, we define a list with process
variables that describe the horizontal positions on all BPMs, and initialize
the interactive display with {\tt  plt.ion()}. Inside the forever-loop, the
BPM positions are read with the {\tt epics.caget\_many()} function and
displayed as a bar graph. The left-hand panel in Figure~\ref{fig:sbpm} shows
a screenshot of the program running. The standard deviation of the beam
positions is shown in the upper left corner of each panel. Instead of
separating the extension {\tt XPOS} with a period, we use a colon, because
the period is a protected character in our default epics installation. 
\par
In much the same way, we can prepare the interface to set the excitations
of the steering magnets with sliders, shown on the right-hand side in
Figure~\ref{fig:sbpm}. The user interface is generated automatically from
the beamline description file and uses {\tt tkinter} commands. For
each corrector, a callback function is defined in which the slider position
is linked to a call to {\tt epics.caput}, a label with the device name, and
a slider to set the magnet. Additionally, {\tt caget} obtains the value and
presets the slider value to the current one. A corresponding interface is
available for the vertical steering magnets, for quadrupoles, for solenoids,
and for the Wien-filter angles. We refer to the github~\cite{MYGH}
for the Matlab code to generate the interface and the python code.
\par
The interface for steering magnets and BPM shown in Figure~\ref{fig:sbpm}
invites to basic {\em gamification} training of new staff. After misaligning
quadrupoles and solenoid magnets with {\tt DT:MISALIGN 0.3}, the student
is tasked to minimize the rms beam positions by adjusting the horizontal
and vertical corrector magnets within a given time frame. The colleague
who reaches a previously set threshold in the shortest time, wins. Of
course, the same  exercise could be done within Matlab, but using the
interfaces used in normal operation will make it easier for the student
to gain familiarity with the CEBAF injector.  
\par
Instead of steering the beamline by hand, we can use the twin to develop
an orbit correction system by first producing the response matrix between
correctors and BPM. To do so we prepare a Python script to scan one
corrector at a time and recording the changes of the beam positions
in both transverse planes in order to accommodate cross-plane coupling.
This gives us a matrix {\tt dBPM(i)/dCOR(j)} for BPMs labeled by {\tt i}
and correctors labeled by {\tt j}. Inverting this matrix tells us the
required changes in the corrector excitation to undo the measured beam
positions. The scripts {\tt make\_response\_matrix.py} and
{\tt correct\_orbit.py} on github~\cite{MYGH} illustrate the process.
\par
\begin{figure}[tb]
\begin{center}
  \includegraphics[width=0.9\textwidth]{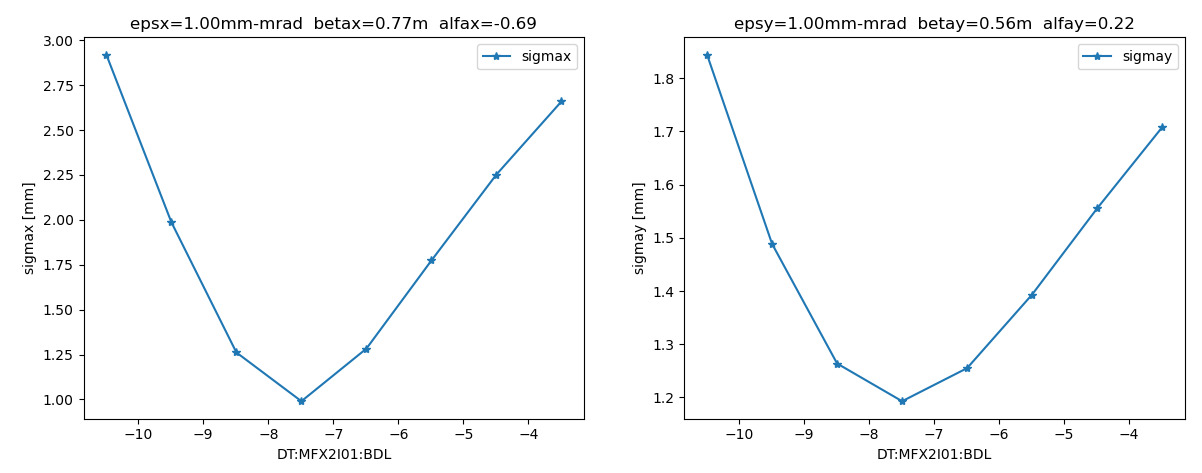}
\end{center}
\caption{\label{fig:qscan}Horizontal (left) and vertical (right) beam sizes
  while scanning the excitation $k_1l$ of the first solenoid doublet. The
  Twiss parameters are reported in the title.}
\end{figure}
The transverse distribution of the beam is continuously updated on the 
screens. Saving the images to disk and accessing the beam sizes on the
selected screen is accomplished by the EPICS process variables
{\tt DT:SCREEN} and {\tt DT:SIGMAS} as described before. Using this
interface it is straightforward to write a script to simulate a
quadrupole scan to determine the Twiss parameters of the beam. We
just scan one or several quadrupoles or solenoids and simultaneously
record the beam sizes from the screen. Extracting Twiss parameters
from the data is well-documented in the literature~\cite{VZAPB} and
relies on knowing the transfer-matrix elements between the magnet
and the screen. Therefore we added retrieval of transfer matrix
elements from the twin. {\tt caput DT:TM:FROM MFX2I01\_US} sets the
start point for the calculation to immediately upstream of the
element named {\tt MFX2I01\_US}. The end point at the screen is set
with {\tt caput DT:TM:TO ITV1I03}. This command also returns the
current $4\times4$ transfer-matrix to EPICS so that its 16 matrix
elements can be retrieved with {\tt caget DT:TM}. This interface
provides the basic functionality to make a quad scan in the
digital twin, all controlled from EPICS. Figure~\ref{fig:qscan}
illustrates the output of the Python script {\tt quad\_scan.py}
from~\cite{MYGH}. The Twiss parameters reported in the title bar
agree with the values from the simulation.
\par
\begin{figure}[tb]
\begin{center}
  \includegraphics[width=0.8\textwidth]{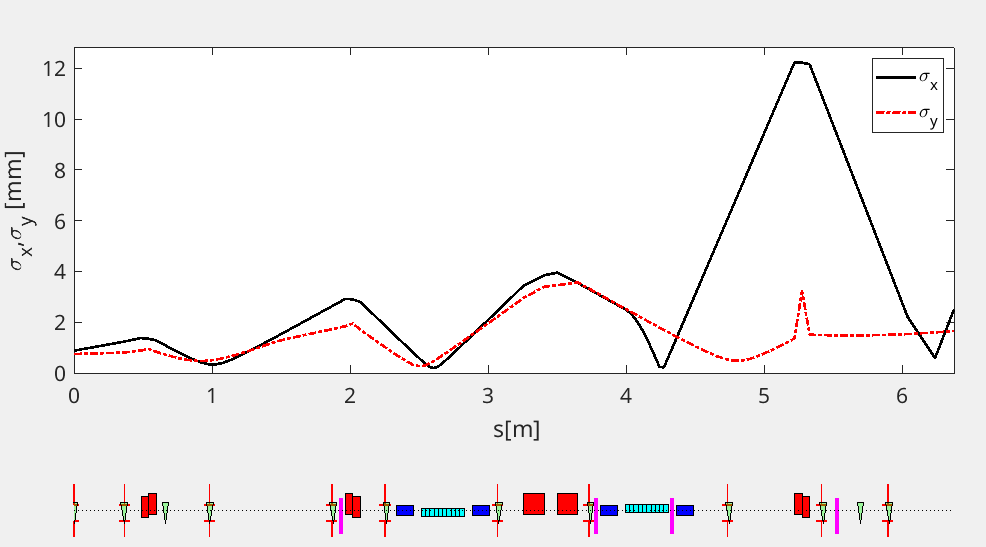}
\end{center}
\caption{\label{fig:hwien}The horizontal Wien filter is adjusted
  to provide $90^o$ spin rotation, which also has  a profound
  influence on the horizontal beam size.}
\end{figure}
The transverse part of the transfer matrices for the two Wien filters
are included in the twin and allow us to explore their influence on
the beam sizes. They very strongly focus the beam in their magnetic
bend plane and have a significant impact on the beam optics.
Figure~\ref{fig:hwien} shows the beam sizes when adjusting the
horizontal Wien filter to rotate the spin by $90^o$. We observe that
it over-focuses the horizontal beam size to a point inside the Wien
(at about $s=4.2\,$m) filter such that it diverges rapidly and causes
the horizontal beam size to exceed 10\,mm in the last solenoid. This
is particularly apparent when comparing with the upper panel in
Figure~\ref{fig:dt1}. Using the twin, it is now possible to
explore compensation algorithms and then use the same interface to
apply them to the real machine.
\par
Exposing all magnets and diagnostic devices via an EPICS interface
might provide an attractive playground to explore machine learning
methods, such that they can later be deployed using the same interface
to work on the real accelerator.
\section{Conclusions}
We developed a basic framework to interface a beam optics simulation
code, written in Matlab, to an EPICS control system. The key component
is the text-based protocol that is very easy to decode in Matlab. It
is equally easy to configure EPICS with the help of the StreamDevice
module. Once this system is running, we can use any EPICS interface,
such as {\tt caget} and {\tt caput} on the command line but also
the epics module for python to access all process variables.
\par
We illustrated the flexibility with a number of python applications,
for example to read the BPM, to adjust magnets, to read screens, and
to retrieve transfer-matrix elements from the twin. Given these basic
applications, more complex ones are easily implemented and tested. A
main point is that developing these applications do not affect the
real accelerator and therefore provide a safe sandbox to develop
new applications.
\par
The system is easily extendable by introducing new commands to listen
to on Matlab and writing the corresponding protocol and database files
for EPICS. Since many elements are very similar, such as the steering
magnets, writing the EPICS files is easily automated. In this way, a
moderately complete set of applications can be generated automatically.
\par
Adapting  this framework to another accelerator is straightforward. All
we have to do is to prepare the beamline description file that conforms
with the seven column and one line per element format described in the
beginning of Section~\ref{sec:twin} which is an extension of the format
used in~\cite{VZAPB}. Adding new types of elements, such as the Wien
filters, is done by preparing functions returning the transfer matrices,
assigning a code to the new element, and adding it to the function
named {\tt calcmat()} that does most of the beam optics calculations.
\par
Using this framework to control other simulation programs should also
be possible. It requires embedding the simulation engine in an infinite
loop that reads from a network socket, decodes the received command,
changes internal parameters of the simulation accordingly, and
then runs the simulation to produce the desired output. 
\section*{Acknowledgments}
We thank the NSF-sponsored REU program that makes the summer student
program at Jefferson Lab possible. One of the authors (JD) was supported
by the U.S. National Science Foundation Research Experience for Undergraduates
at Old Dominion University and Jefferson Lab Grant No. 2348822. Moreover,
this material is based upon work supported by the U.S. Department of Energy,
Office of Science, Office of Nuclear Physics under Contract No. 89243126CSC000213.
%
%
\bibliographystyle{plain}

\end{document}